\documentstyle{article}
\begin{document}

%                          Title
\begin{center}
{\Large \bf Renormalization prescriptions and bootstrap \\
in effective theories} \\

\vspace{4mm}

%                      author/address
%Author's name\\
%Author's institute \\
% and address\\
\underline{V.~Vereshagin}$^{a}$, 
K.~Semenov-Tian-Shansky$^{a,b}$  and
A.~Vereshagin$^{a,c}$ \\
$^a$St.-Petersburg State University,
$^b$Universit\'e  de Li\`ege au Sart Tilman and
$^c$University of Bergen\\
vvv@av2467.spb.edu\\

\end{center}

%                        Abstract
\begin{abstract}
We discuss the peculiar features of the renormalization procedure in
the case of infinite-component effective theory. It is shown that in
the case of physically interesting theories (namely, those leading to
the amplitudes with asymptotic behavior governed by known Regge
intercepts) the full system of required renormalization prescriptions
only contains those needed to fix the minimal counterterm vertices
with one, two, three and (in some cases) four lines. There is no
necessity in imposing the prescriptions for n-leg counterterm
vertices with
$n > 4$.
Moreover, the prescriptions for
$n \leq 4$
cannot be taken arbitrary: an infinite system of bootstrap constraints
must be taken into account. The general method allowing one to write
down the explicit form of this system is explained and illustrated by
the example of elastic scattering process.
\end{abstract}

%%%%%%%%%%%%%%%%%%%%%%%%%%%%%%%%%%%%%%%%%%%%%%%%%%%%%%%%%%%%%%%%%%%%%%
\section{Introduction}
\label{sec-intro}
\mbox{}

In this talk we give a brief description of the mathematical scheme
that allows one to put in order an infinite set of renormalization
prescriptions (RP's) needed to fix the physical content of the
{\em localizable effective scattering theory}%
\footnote{The preliminary discussion of the problem and definitions
of the terms can be found in
\cite{AVVV1}--\cite{AVVV2}, see also our talks
\cite{HSQCDKS}--\cite{HSQCDVV} given at HSQCD'2004 and two another
talks
\cite{QFTHEP} at this conference.
}
of strong interactions. For simplicity we only
consider the non-strange sector. This means that all the true
asymptotic states are constructed from pions and nucleons and the free
Hamiltonian is solely defined by the operators of these fields (we
rely upon the intrinsically quantum construction of the field theory
described in
\cite{WeinMONO}).
An infinite set of auxiliary resonance fields only appears in the
{\em extended perturbation scheme}
which we use to assign rigorous perturbative meaning to the formal
Dyson's series of the initial effective theory. It is this scheme
which we discuss below. Here we have no space for detailed discussion
--- it will be published elsewhere
\cite{KSAVVV2}.

%%%%%%%%%%%%%%%%%%%%%%%%%%%%%%%%%%%%%%%%%%%%%%%%%%%%%%%%%%%%%%%%%%%%70

\section{The Cauchy formula in hyper-layers}
\label{sec-DR1}
\mbox{}

In this Section we remind the main features of the mathematical tool
that turns out especially useful for constructing the explicit form of
the system of bootstrap conditions.

\begin{figure}[ht]
%\begin{center}
\begin{picture}(200,140)

 %Axes
\put(85,60){\vector(1,0){235}}
 \put(200,0){\vector(0,1){140}}

 %Axis names
\put(205,130){{\small {\it Im}} Z}
            %{{\small $\mathcal{I}$mz  }}   this works perfectly!
 \put(305,48){{\small {\it Re}} Z}
            %{{\small $\mathcal{R}ez$  }} %this command works a bit 
                                          %strange!

 %Contour C_{m,n}
\put(230,115){{\small $C_{m,n}$  }}
 \put(200,60){\oval(180,100)[t]}
  \put(200,60){\oval(180,100)[b]}
%   \put(270,60){\oval(40,100)[br]}    %{\oval(40,100)[r]}
%    \put(130,60){\oval(40,100)[tl]}   %{\oval(40,100)[l]}
%     \put(180,10){\line(1,0){40}}
%      \put(180,110){\line(1,0){20}}

\put(130,110){\line(1,0){45}}
 \put(270,10){\line(-1,0){45}}
% \put(270,10){\line(-1,0){15}}

\put(220,10){\line(1,0){15}}
 \put(165,110){\line(1,0){15}}

 %Poles (inside), cuts and contours C_k
\put(135,60){\circle*{4}}
 \put(130,50){{\tiny $s_{-m}$}}
  \put(135,60){\oval(10,10)[b]}
   \multiput(135,58)(0,6){11}{\line(0,1){4}}
    \put(130,60){\line(0,1){50}}
     \put(140,60){\line(0,1){50}}

\put(143,90){{\tiny $C_{-m}$}}
\put(130,125){{\tiny Cut}}

\put(175,60){\circle*{4}}
 \put(170,50){{\tiny $s_{-1}$}}
  \put(175,60){\oval(10,10)[b]}
   \multiput(175,58)(0,6){11}{\line(0,1){4}}
    \put(170,60){\line(0,1){50}}
     \put(180,60){\line(0,1){50}}

\put(183,90){{\tiny $C_{-1}$}}
\put(170,125){{\tiny Cut}}

\put(225,60){\circle*{4}}
 \put(225,60){\oval(10,10)[t]}
  \put(223,68){{\tiny $s_1$}}
   \multiput(225,62)(0,-6){11}{\line(0,-1){4}}
    \put(220,60){\line(0,-1){50}}
     \put(230,60){\line(0,-1){50}}

\put(233,25){{\tiny $C_1$}}

\put(233,68){{\small \,. . . .}}
 \put(230,12){{\small \,. . . .}}

\put(140,50){{\small \,. . .}}
 \put(140,106){{\small \,. . . .}}

\put(265,60){\circle*{4}}
 \put(263,68){{\tiny $s_n$}}
  \put(265,60){\oval(10,10)[t]}
   \multiput(265,62)(0,-6){11}{\line(0,-1){4}}
    \put(260,60){\line(0,-1){50}}
     \put(270,60){\line(0,-1){50}}

\put(273,25){{\tiny $C_n$}}

 %Poles (outside)
\put(95,60){\circle*{4}}
 \put(85,68){{\tiny $s_{-m-1}$}}

\put(305,60){\circle*{4}}
 \put(300,68){{\tiny $s_{n+1}$}}
\end{picture}
\caption{\em  System of contours on the complex-$z$ plane.
\label{SysofCon}}
%\end{center}
\end{figure}
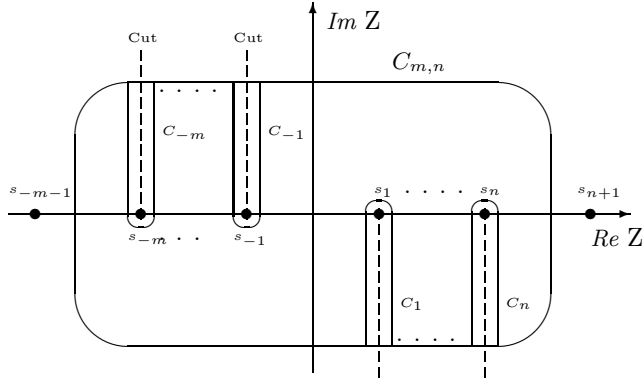

Consider the function
$f(z,{\bf x})$
analytic in complex variable
$z$
and smoothly depending on a set of parameters
${\bf x} \equiv  \{ x _i \}$.
Let us further suppose that when
${\bf x} \in D$
(here
$D$
is a small domain in the space of parameters) this function has only a
finite number of singular points in every finite domain of the
complex-$z$ plane. On
Fig.~\ref{SysofCon}
it is shown the geography of singular points
$s_k$
$(k = 0, \pm 1, \pm 2, \ldots)$
typical for the finite loop order scattering amplitudes in quantum
field theory. Note that the cuts are drawn in unconventional way ---
this is done for the sake of convenience. If the point
$s_k$
corresponds to the pole type singularity there is no need in cut, but
its presence makes no influence on the results discussed below.

Let us recall the definition of the polynomial boundedness property
adjusted
\cite{AVVV1}
for the case of many variables. Consider the system of closed embedded
contours
$C(i) \equiv  C_{-m_i,n_i}$
(see Fig.~\ref{SysofCon})
%%%%%%%%%%%%%%%%%%%%%%%%%%%%%%%%%%%%%%%%%%%%%%%%%%%%%%%%%%%%%%%%%%%%70
%%%%%%%%%%%%%%%%%%%%%%%%%%%%%%%%%%%%%%%%%%%%%%%%%%%%%%%%%%%%%%%%%%%%70
such that
$m_i \geq m_j$
and
$n_i \geq n_j$
when
$i \geq j$
(both left and right singular points are enumerated in order of
increasing modulo). We say that the function
$f(z,{\bf x})$ is
$N$-bounded in the hyper-layer
$B_x\{z \in {\bf C},\, {\bf x} \in D \}$
if there is an infinite system of contours
$C(i)$ $(i=1,2,\ldots)$ and an integer
$N$ such that, when
$i \rightarrow \infty$,
\begin{equation}
\max_{{\bf x} \in D;\ z \in C(i)}
{\left|\frac{f(z,{\bf x})}{z^{N+1}}\right|}
\longrightarrow 0\ .
\label{3.1}
\end{equation}
The
{\em minimal}
$N$ (possibly, negative) providing the correctness of the uniform (in
$x$) estimate
(\ref{3.1}) we call the degree of bounding polynomial in
$B_x$.

Let us now make use of the famous Cauchy's integral formula for the
function
$f(z,{\bf x})/z^{N+1}$.
Consider the closed contour (see
Fig.~\ref{SysofCon})
consisting of
$C(i)$
(except small segments crossing the cuts) and the corresponding
parts of the contours
$C_{-p}$, $C_{q}$ $(p=1,2,...,m_i;\; q=1,2,...,n_i;)$
surrounding cuts%
\footnote{
We assume that
$f$
is regular at the origin. However
$f(z,{\bf x})/z^{N+1}$
may have a singularity there and to apply the Cauchy theorem one
should add a circle around the origin to the contour of integration.
It is this part of the contour that gives the first sum in the right
side of
Eq.~(\ref{3.2}).
}.
Taking the limit
$i \to \infty$
and keeping in mind
(\ref{3.1})
one obtains
\begin{equation}
f(z, {\bf x}) =
\sum \limits_{k=0}^{N}
\frac{1}{k!} f^{(k)}(0, {\bf x}) z^k +
\frac{z^{N+1}}{2 \pi i}
\sum \limits_{k= - \infty}^{\infty}\;
\int \limits_{C_k}^{}
\frac{f(\xi , {\bf x})}{{\xi}^{N+1} (\xi - z)}\; dz \; ,
\label{3.2}
\end{equation}
where the last sum is done in order of increasing modulo of the
singularities
$s_k$ which the contours
$C_k$ are drawn around. The relation
(\ref{3.2}) provides a mathematically correct form of the result. In
the case when the number of singular points is infinite, every contour
integral on the right side should be considered as a single term of
the series. The mentioned above order of summation provides a
guarantee of the uniform (in
$z$
and
${\bf x}$) convergence of the series.

The formula
(\ref{3.2})
plays the key role in the renormalization programme discussed below.

%%%%%%%%%%%%%%%%%%%%%%%%%%%%%%%%%%%%%%%%%%%%%%%%%%%%%%%%%%%%%%%%%%%%70
\section{Tentative consideration}
\label{sec-RP1}
\mbox{}

As noted in
\cite{POMI},
the problem of ordering the
full set of renormalization prescriptions needed to fix the physical
content of the effective scattering theory should be considered in
terms of the resultant parameters
\cite{AVVV2}.
The simple example below explains the main idea of our renormalization
procedure and shows the source of the bootstrap conditions.

Let us consider the case of elastic scattering process
\begin{equation}
a(p_1) + b(k_1) \to a(p_2) + b(k_2)\ .
\label{4.1}
\end{equation}
For simplicity, we consider both
$a$ and $b$
particles to be spinless. This considerably simplifies the purely
technical details without changing the logical line of the analysis.

Along with the conventional kinematical variables
$s = (k_1 + p_1)^2$,
$t = (k_1 - k_2)^2$,
$u = (k_1 - p_2)^2$
we introduce three equivalent pairs of independent ones:
\begin{equation}
(x, {\nu}_x), \quad x=s,t,u; \quad {\rm where} \;\;
{\nu}_s = u - t,\;
{\nu}_t = s - u,\;
{\nu}_u = t - s.
\label{4.2}
\end{equation}
The pair
$(x, {\nu}_x)$
provides a natural coordinate system in 3-dimensional (one complex and
one real coordinate) layer
$B_x\{{\nu}_x \in {\bf C};\  x \in (a,b)\in {\bf R} \}$.

Let us now suppose that in
$B_t\{{\nu}_t \in {\bf C};\; t \in {\bf R},\; t \sim 0 \}$
the full (non-perturbative) amplitude of the process
(\ref{4.1})
is described by the 0-bounded function
$f({\nu}_t, t)$
($N_t = 0$).
According to the uniformity principle%
\footnote{The precise formulation of this principle is given in
\cite{HSQCDVV}.
It looks as follows.
{\em The degrees of the bounding polynomials which specify the
asymptotic properties of individual items of the loop series expansion
must be taken equal to the degree of the polynomial specifying the
asymptotics of the full (non-perturbative) amplitude of the process
under consideration}.
},
we have to construct the perturbation series
$$
f({\nu}_t, t) = \sum \limits_{l=0}^{\infty} f_l({\nu}_t, t)
$$
in such a way that the full sum
$f_l(...)$
of the
$l$-th
loop order graphs must also present the 0-bounded function in
$B_t$.
Hence, from the relation
(\ref{3.2})
it follows that in this layer
\begin{equation}
f_l({\nu}_t,\; t ) = f_l(0,t) +
\frac{{\nu}_t}{2 \pi i}
\sum \limits_{k = -\infty}^{\infty}\;
\int \limits_{C_k (t)}^{}
\frac{f_l(\xi , t)}{{\xi} (\xi - {\nu}_t)}\; d{\xi}\ \ .
\label{4.3}
\end{equation}
Here the notation
$C_k (t)$
is used to stress that we deal with singularities (and, hence, with
cuts) in the complex-${\nu}_t$ plane; the variable
$t$
should be considered as a parameter.

We suppose that all the numerical parameters, needed to fix the finite
(renormalized) amplitudes of the previous loop orders, are known and
one only needs to carry out the renormalization of the
$l$-th
order graphs. To do this let us assume (this assumption is justified
below) for the moment  that the infinite sum of integrals in
(\ref{4.3})
solely depends on the parameters already fixed on the previous steps
of renormalization procedure. Thus it only remains to fix the function
$f_l(0, t)$ --- then 
(\ref{4.3}) will give the complete renormalized expression for the
$l$-th order contribution in
$B_t$.
It is sufficient to fix the values of the coefficients in power
series expansion of
$f_l(0, t)$.
This can be done with the help of self-consistency requirement.

To make use of this requirement we need to consider the
cross-conjugated process
\begin{equation}
a(p_1) + \overline{a}(-p_2) \to  \overline{b}(-k_1) +b(k_2)\ .
\label{4.4}
\end{equation}
Let us suppose that in
$B_u\{{\nu}_u \in {\bf C};\; u \in {\bf R},\; u \sim 0 \}$
it is described by the (-1)-bounded
$(N_u = -1)$
amplitude
$$
\phi ({\nu}_u,\; u) =
\sum \limits_{l=0}^{\infty} {\phi}_l({\nu}_u,\; u)\; .
$$
The uniformity principle tells us that every function
${\phi}_l({\nu}_u,\; u)$,
in turn, must be (-1)-bounded in
$B_u$.
Hence in this layer
\begin{equation} {\phi}_l({\nu}_u,\; u ) = \frac{1}{2 \pi i}
\sum \limits_{k = -\infty}^{\infty}\;
\int \limits_{C_k (u) }^{}
\frac{{\phi}_l(\xi , u)}{(\xi - {\nu}_u)}\; d{\xi}\ \ .
\label{4.5}
\end{equation}
Again, it is assumed (and proved below) that the sum of integrals
in the right side only depends on the parameters already fixed on the
previous steps of the renormalization procedure.

Remembering now that both expressions
(\ref{4.3}) and
(\ref{4.5}) follow from the same infinite sum of
$l$-loop graphs we conclude that in the intersection domain
$D_s \equiv B_t \cap B_u$ they must coincide with one another:
\begin{equation}
f_l(0,t) +
\frac{{\nu}_t}{2 \pi i}
\sum \limits_{k = -\infty}^{\infty}\;
\int \limits_{C_k (t)}^{}
\frac{f_l(\xi , t)}{{\xi} (\xi - {\nu}_t)}\; d{\xi} =
\frac{1}{2 \pi i}
\sum \limits_{k = -\infty}^{\infty}\;
\int \limits_{C_k (u)}^{}
\frac{{\phi}_l(\xi , u)}{(\xi - {\nu}_u)}\; d{\xi}\ \ ,
\label{4.6}
\end{equation}
which means that in
$D_s$
\begin{equation}
f_l(0,t) = -\frac{{\nu}_t}{2 \pi i}
\sum\limits_{k =-\infty}^{\infty}\;
\int \limits_{C_k (t)}^{}
\frac{f_l(\xi , t)}{{\xi} (\xi - {\nu}_t)}\; d{\xi} +
\frac{1}{2 \pi i}
\sum \limits_{k = -\infty}^{\infty}\;
\int \limits_{C_k (u)}^{}
\frac{{\phi}_l(\xi , u)}{(\xi - {\nu}_u)}\; d{\xi}\
\equiv {\Psi}^{(0,-1)} (t,u)\ .
\label{4.7}
\end{equation}

The relation
(\ref{4.7}) only makes sense in
$D_s$. It is not difficult to construct two more relations of this
kind, one of them being valid in
$D_t \equiv B_u \cap B_s\,$, and other one --- in
$D_u \equiv B_s \cap B_t\,$. These relations play a key role in our
approach because they provide us with a source of an infinite system
of bootstrap conditions. To explain this point, let us consider
(\ref{4.7}) in more detail and make two statements.

First one:
{\em despite of the fact that
(\ref{4.7})
only makes sense in
$D_s$,
it allows one to express all the coefficients
$c_k(l)$
of the power series expansion
$$
f_l(0,t) = \sum \limits_{k=0}^{\infty}c_k(l)t^k
$$ in terms of the parameters which, by suggestion, have been fixed on
the previous steps of renormalization procedure}. The set of those
coefficients completely defines this function everywhere in
$B_t$. When translated to the language of Feynman rules, this means
that in our model example there is no necessity in attracting special
renormalization prescriptions fixing the finite part of the four-leg
counterterms. Instead, the relation
(\ref{4.7})
should be treated as that generating the relevant RP's iteratively ---
step by step. In what follows we call this --- generating --- part of
self-consistency equations as
{\em bootstrap conditions of the first kind}.

Second:
{\em the relation
(\ref{4.7})
strongly restricts the allowed values of the parameters which are
assumed to be fixed on the previous stages}.
To show this it is sufficient to note that
$f_l(0,t)$
only depends on the variable
$t$
while the function
${\Psi}^{(0,-1)} (t,u)$
formally depends on both variables
$t$ and $u$.
Thus we are forced to require the dependence on
$u$
to be fictitious. It is this requirement which provides us with an
additional infinite set of restrictions for the resultant parameters.
We call these restrictions as
{\em the bootstrap conditions of the second kind}.

The proof of both statements is simple. The domain
$D_s\{ t \sim 0,\; u \sim 0 \}$
contains the point
$(t=0,\; u=0)$
which can be taken as the origin of the local coordinate system. Using
the definitions
(\ref{4.2})
and expanding both sides of
(\ref{4.7})
in power series in
$t$
and
$u$
we obtain the full set of Taylor coefficients
\begin{equation}
c_k(l) = \frac{1}{n!}\  f_l^{(k)}(0,0) =
\frac{1}{n!}\  \frac{{\partial}^k}{{\partial t}^k}\
{\Psi}_l^{(0,-1)}(t=0,u=0),\ \ \ \ \ \ \ \ \ (k=0,1,\ldots)\; .
\label{4.8}
\end{equation}
This is quite sufficient for fixing
$f_l(0,t)$
everywhere in
$B_t$.
Further, from
(\ref{4.7})
it follows an infinite system of the bootstrap conditions of the
second kind:
\begin{equation}
\frac{{\partial}^k}{{\partial t}^k}
\frac{{\partial}^{m+1}}{{\partial u}^{m+1}}
 {\Psi}_l^{(0,-1)}(t=0,u=0)\ = 0,\ \ \ \ \ \ \ (k,m=0,1,\ldots)\; ;
\label{4.9}
\end{equation}
they restrict the allowed values of the parameters fixed on the
previous steps of renormalization procedure.

Thus we see that the system of bootstrap equations%
\footnote{
This is not the full system: it only mirrors the self-consistency
requirements in certain domains of the complex space of relevant
kinematical variables!
}
is naturally divided into two subsystems. The bootstrap conditions of
the first kind just allow one to express the resultant parameters of
higher levels in terms of the lower level parameters which, by
condition, already have been expressed in terms of the
{\em fundamental observables}%
\footnote{
The parameters appearing in the right sides of RP's.
}
on the previous steps. In other words, they provide a possibility to
express the higher level parameters in terms of observable quantities.
This subsystem does not restrict the admissible values of the latter
quantities.

In contrast, the subsystem consisting of the bootstrap conditions of
the second kind does impose extremely strong limitations on the
allowed values of the physical (observable!) parameters of effective
scattering theory. In fact, it provides us with the system of physical
predictions which --- at least, in principle --- can be verified
experimentally.

To make our analysis complete it remains to discuss two key points.
First, we need to explain the above-made suggestions concerning the
degrees of bounding polynomials. Second, it is necessary to show that
all the contour integrals in
(\ref{4.3}) and (\ref{4.5})
only depend on the parameters already fixed on the previous stages of
the renormalization procedure. The first point can be easily
explained: we choose the bounding polynomial degrees in accordance
with known data on Regge intercepts. The proof of the second one is
based on the relation
(\ref{3.2})
and on the results of
\cite{AVVV2}.

First of all we need to review the main stages of the renormalization
procedure. As usually, we discuss this point in terms
of 1PI (one-particle irreducible) graphs%
\footnote{
In
\cite{AVVV2}
we considered tadpoles (1-leg graphs) attached to a given vertex on
the same footing as self-closed lines. Here, however, we consider
tadpoles as independent elements of Feynman rules for constructing
graphs in terms of resultant parameters. This allows us to avoid
problems connected with the definition of one-particle
irreducibility.
} 
(see, e.g.,
\cite{Renorm}, \cite{Peskin}).
Again, we consider first the elastic two body scattering
(\ref{4.1}).
As follows from the analysis in
\cite{AVVV2},
the reduction procedure does not change the structure of singularities
of a given graph; it only allows one to re-express this graph in terms
of minimal parameters of various levels. This relates also to the full
sum of graphs of a given loop order. The reduction procedure just
converts it into another sum of graphs, each one being written in
terms of resultant parameters. We always imply that the reduction is
done and work with these latter parameters.

Let us begin from the consideration of one loop contributions.
In this case one deals with the graphs solely constructed from the
minimal propagators and resultant 1-, 2-, 3- and 4-leg vertices with
the level indices
$l=0,1$.
First, there are graphs with one
{\em explicit loop}:
1) Those containing one minimal tadpole (1-leg) insertion, 
2) Those with one minimal self-energy type (2-leg) insertion, 
3) Graphs with one-loop minimal triple vertex and, at last, 
4) One-particle irreducible one-loop graphs. All these graphs only 
depend on the parameters of the lowest level
$l=0$.
Second, there are graphs with one
{\em implicit loop}:
those containing the completely reduced 1-, 2-, 3- and 4-leg vertices
and the same graphs with insertions of 1-loop
counterterms for 1-, 2-, 3- and 4-leg minimal effective vertices. All
the parameters of the highest level
$l=1$
are concentrated in the graphs of this latter kind. As usually, to fix
the one-loop order counterterms for 4-point amplitude, one shall first
perform the one-loop renormalization of the resultant vertices with
one, two and three legs (tadpole, self-energy and triple vertex). To 
fix the remaining 4-leg counterterm one needs to formulate the 
corresponding renormalization prescription.

Clearly, this conclusion can be easily generalized for the case of
inelastic processes involving an arbitrary number of particles. This
is precisely the situation known from the conventional renormalization
theory. The only difference is that in the case of effective
scattering theory with unstable particles the wave function
renormalization process has
certain peculiar features%
\footnote{This point is briefly discussed in
\cite{AVVV2};
the detailed analysis will be published elsewhere.}.
Besides, in this latter case the full number of counterterms, needed
to construct the finite expressions for all one-loop
$S$-matrix
elements, is infinite.

The generalization of the above reasoning for the case of arbitrary
loop order is straightforward. Namely, the procedure of
renormalization of the
$L$-loop
contribution to the amplitude of the process involving
$N$
($N=4,5, \ldots$)
particles consists of
$L$
stages. Every
$l$-th
stage
$(l=1,2,\ldots ,(L-1))$
presents a certain (depending on
$N$ and $l$)
number of steps each of which corresponds to the renormalization of
$l$-loop
$S$-matrix
elements with fixed number
$n$
of external lines
$(n = 1,2,\ldots , n_{max}(N,l))$.
The last ---
$L$-th ---
stage consists of
$(N-1)$
preliminary steps (renormalization of the
$L$-loop
$n$-leg
graphs with
$1 \leq n < N$)
and one final step --- fixing the
$L$-th
order
$N$-leg
counterterm vertices.

Now we can turn to a consideration of contour integrals in
(\ref{4.3}) and (\ref{4.5}).
It is implied that all the previous steps already have been done and
we only need to fix the
$l$-th
level coefficients in the series expansion of 4-leg vertex.

It is clear%
\footnote{
This point is justified by the
{\em summability} 
principle formulated in
\cite{HSQCDVV}.
}
that the only contributions to the contour integrals in question
follow from those graphs which have at least one internal line
(otherwise the graph does not contain a singularity). It is easy to
show that the graphs with internal lines may only depend on the 
parameters of the resultant vertices of lower levels and on the
$l$-th level resultant parameters describing the vertices with 1, 2,
or 3 legs. By condition, all those parameters have been already fixed
on the previous steps of renormalization procedure. All this means 
that, indeed, the contour integrals in
(\ref{4.3}) and (\ref{4.5}) should be considered as known functions. 
This completes our proof.

To summarize: in our example, as far as we consider all the resultant
parameters of the previous levels
$l^{\prime} = 1,2,\ldots ,(l-1)$
being fixed, one does not need to attract any new RP's in addition to
those fixing the
$l$-th level resultant vertices with
$n=1,2,3$ legs; the 4-leg counterterms of the
$l$-th loop level automatically become fixed by the bootstrap 
conditions of the first kind.

The above analysis offers a hint about the structure of the full
system of requirements sufficient for performing the renormalization
procedure in effective scattering theory.

%%%%%%%%%%%%%%%%%%%%%%%%%%%%%%%%%%%%%%%%%%%%%%%%%%%%%%%%%%%%%%%%%%%%70
\section{Renormalization programme}
\label{sec-RP3}
\mbox{}

The model example considered in the previous section allows us to make
the following statement. In those cases when it is possible to
point out two intersecting hyper-layers
$B_x$ and 
$B_y$ such that in one of them (say, in
$B_x$) the amplitude of a given process
$2 \rightarrow 2$ is presented by the 
$(-1)$-bounded function of the corresponding complex variable (and, 
possibly, of several parameters) there is no need in
formulating the RP's for 4-leg amplitudes: the summability principle
provides us with a tool for generating those prescriptions order by
order. This conclusion remains in force also if in another layer 
$B_y$ the degree of the relevant bounding polynomial is greater than
one
$N_y \geq 1$.
As to the 1-, 2- and 3-leg vertices (which all are just constants in
terms of the resultant parameters), one does need to formulate the
relevant RP's, but this cannot be done arbitrarily --- the bootstrap
requirements must be taken into account.

It is well known that the amplitudes of inelastic processes involving
$n > 4$ particles decrease with energy, at least in the physical area
of the other relevant variables. Thus it looks natural to suggest that
in corresponding hyper-layers these amplitudes can be described with 
the help of at most
$(-1)$-bounded functions of one complex energy-like
variable (and several parameters). Also, it is always possible to
choose the variables in such a way that the domains of mutual
intersections of
every two hyper-layers are non-empty%
\footnote{
This follows from the fact that the number of pair energies is much
larger then that of independent kinematic variables.
}.
Therefore, from the above analysis it follows that the system of RP's
needed to fix the physically interesting effective scattering theory
only contains those prescriptions which fix the 1-, 2-, 3- and,
possibly, 4-leg amplitudes%
\footnote{
We would have no need in RP's even for 4-leg amplitudes if the
experimental information on hyper-layers where they decrease with
energy were more complete.
}.

We would like to recall that we are interested in constructing the
renormalization procedure which could provide a possibility to manage
the effective hadron scattering theory. For simplicity, we only
consider here the non-strange sector%
\footnote{
The situation in the strange sector is discussed in
\cite{HSQCDKS}.
}.
This means that the only stable particles in our case are pions and
nucleons. As known, the high-energy behavior of the amplitudes of
elastic pion-pion, pion-nucleon and nucleon-nucleon scattering
processes is governed by the Regge asymptotic law. It is not difficult
to check that every one of these processes is described by the
amplitude (more precisely, by several scalar formfactors)
characterized by the 
{\em negative} degree of bounding polynomial, at least in one of the
three cross-conjugated channels. That is why the analysis in
Sec.~\ref{sec-RP1}
is relevant, and we conclude
that 4-leg graphs with external lines corresponding to pions and
nucleons do not require formulating RP's.

Now we need to consider the 4-leg graphs with at least one external 
line corresponding to unstable particle. Here the situation looks more
complicated owing to the absence of direct
experimental information about the processes with unstable hadrons.
In other words, the problem of choice of the degrees of relevant
bounding polynomials is to a large measure nothing but a matter of
postulate. The only way to check the correctness of the choice
is to construct the corresponding bootstrap relations and
compare them with existing data on resonance parameters. This work
is in progress now. Here, however, we consider the relatively simple
situation when all the 4-leg amplitudes of the processes involving
unstable particles decrease with energy, at least, at sufficiently
small values of the momentum transfer%
\footnote{
Surely, this is just a model suggestion which, however,
seems us quite reasonable. One of the arguments in its favor follows
from the fact that unstable particles cannot appear in true asymptotic
states.
}.
As we already mentioned, the same relates also to the amplitudes of
the processes with
$n > 4$
external particles. All this means that, according to the results
discussed above,
the only RP's needed to fix the physical content of the effective
theory under consideration are those restricting the allowed values of
the coupling constants at 1-, 2- and 3-leg minimal counterterm
vertices.

We stress that this conclusion is based on the analysis of the
bootstrap equations only valid in the intersection domain of two
layers
$B_x\{ x \in {\bf R},\ x \sim 0;\ {\nu}_x \in {\bf C}\} $
and
$B_y\{ y \in {\bf R},\ y \sim 0;\ {\nu}_y \in {\bf C}\} $,
each of which contains small vicinity of the origin of the relevant
coordinate system. The reason for this choice of layers is explained
by the existence of experimental information on Regge intercepts. This
choice is, however, justified  from the field-theoretic point of view
\cite{AVVV1}--\cite{AVVV2}. Here are the arguments. A formal way to
construct the
$(L+1)$-th order amplitude of the process
$X \longrightarrow Y$ is to close the external lines of the relevant
$L$-th
order graphs corresponding to the process
$X + p_1 \longrightarrow Y + p_2$
with two additional particles carrying the momenta
$p_1$ (let it be incoming) and
$p_2$ (outgoing). This means that those latter graphs should be
calculated at
$p_1 = p_2 \equiv q$, dotted by the relevant propagator and integrated
over
$q$. To ensure the correctness of this procedure (see
\cite{axioms}), one needs to require the polynomial boundedness (in
$p_1$) of the
$L$-th order amplitude of the process
$X + p_1 \longrightarrow Y + p_2$ at
$x \equiv (p_1 - p_2)^2 = 0$ and, by continuity, in a small vicinity 
of this value. Clearly, this argumentation applies to arbitrary graphs
with
$N \geq 4$ external lines.

Thus we arrive at the following conclusion.
{\em To perform the complete renormalization of the effective
scattering theory with the above-specified asymptotic conditions
it is quite sufficient to attract the system of RP's that fix the
finite parts of counterterm coupling constants at minimal
$1-$, $2-$ and
$3-$leg vertices. Moreover, this system cannot be taken arbitrary
--- the bootstrap constraints must be taken into account.}

We are grateful to V.~Cheianov, H.~Nielsen, S.~Paston, J.~Schechter,
A.~Vasiliev and M.~Vyazovski for stimulating discussions. The work was
supported in part by INTAS (project 587, 2000) and by Ministry of
Education of Russia (Programme
``Universities of Russia'').
The work by A.~Vereshagin was supported by L.~Meltzers H\o yskolefond
(Studentprosjektstipend, 2004).

%%%%%%%%%%%%%%%%%%%%%%%%%%%%%%%%%%%%%%%%%%%%%%%%%%%%%%%%%%%%%%%%%%%%70

%%%%%%%%%%%%%%%%%%%%%%%%%%%%%%%%%%%%%%%%%%%%%%%%%%%%%%%%%%%%%%%%%%%%70
\end{document}